\documentclass[journal,10pt]{IEEEtran}

\usepackage{amsmath,amssymb,amsfonts}
\usepackage{graphicx}
\usepackage{textcomp}
\usepackage{xcolor}
\usepackage{algorithm,algpseudocode}
\usepackage{verbatim}

\begin{document}

\title{Energy-Efficient Vehicular Edge Computing with One-by-one Access Scheme}

\author{Youngsu Jang,~\IEEEmembership{Student Member,~IEEE}, Seongah Jeong,~\IEEEmembership{Member,~IEEE}, and Joonhyuk Kang,~\IEEEmembership{Member,~IEEE}

\vspace{-0.5cm}
\thanks{Youngsu Jang and Joonhyuk Kang are with the School of Electrical Engineering, Korea Advanced Institute of Science and Technology, Daejeon, 34141 Korea (e-mail: jangyoung30@kaist.ac.kr, jhkang@kaist.edu).}
\thanks{Seongah Jeong is with the School of Electronics Engineering, Kyungpook National University, Daegu, 14566 Korea (e-mail: seongah@knu.ac.kr).}
\thanks{Manuscript received XXX, XX, 2022; revised XXX, XX, 2022.}}

\markboth{IEEE Transactions on Vehicular Technology,~Vol.~XX, No.~XX, XXX~2022}
{}

\maketitle

\begin{abstract}
With the advent of ever-growing vehicular applications, vehicular edge computing (VEC) has been a promising solution to augment the computing capacity of future smart vehicles. The ultimate challenge to fulfill the quality of service (QoS) is increasingly prominent with constrained computing and communication resources of vehicles. In this paper, we propose an energy-efficient task offloading strategy for VEC system with one-by-one scheduling mechanism, where only one vehicle wakes up at a time to offload with a road side unit (RSU). The goal of system is to minimize the total energy consumption of vehicles by jointly optimizing user scheduling, offloading ratio and bit allocation within a given mission time. To this end, the non-convex and mixed-integer optimization problem is formulated and solved by adopting Lagrange dual problem, whose superior performances are verified via numerical results, as compared to other benchmark schemes.
\end{abstract}

\vspace{-0.2cm}

\IEEEpeerreviewmaketitle

\vspace{-0.3cm}
\section{Introduction}
With the rapid development of vehicular technology including autonomous driving, future vehicles are expected to play a role of providing various infotainment services to users as well as a simple means of transportation. Services such as voice recognition, autonomous driving, video streaming, and virtual reality/augmented reality (VR/AR) require significant computing resources and strict delay constraints, which might not be processed in on-board vehicles with limited computing and battery resources. Vehicular edge computing (VEC) [1-4] has emerged as an economical and scalable alternative to process offloaded data efficiently while providing improved quality of service (QoS) to vehicular users from anywhere and at any time at reduced costs \cite{1}. \\
\indent In VEC systems, an edge server mounted on a road side unit (RSU) located nearest the vehicles can provide additional computational resources for high-complexity applications, which allows to reduce latency as well as save the energy required for offloading procedure. However, in general, the vehicular communication environment rapidly varies due to the high speed and mobility of vehicles, making it difficult to apply the traditional mobile edge computing (MEC)-based offloading method as it is. Therefore, further researches on efficient task offloading strategies suitable for the VEC systems are needed. \\
\indent With the rapid spread of electric vehicles in recent years, efficient use of the vehicle's limited battery capacity has become very important. Due to the constrained energy budget of vehicles, several studies have been conducted on task offloading in VEC systems to minimize the energy consumption [2-4]. The authors in \cite{2} study the energy-efficient workload offloading problem, and propose a low-complexity distributed solution based on consensus alternating direction method of multipliers, but only a single RSU is considered. In \cite{3}, a novel three-layered system, i.e., vehicular edge cloud computing (VECC), is proposed as a solution to energy conservation and computation augmentation for vehicular computing, and a deep learning-assisted energy-efficient task offloading algorithm is developed in \cite{4}. However, \cite{3} does not consider the partial offloading to offload the part of the task, and \cite{4} only consider the user association without considering multiple access scheme, which can be further improved.\\
\indent In this paper, we propose an energy-efficient task offloading strategy in VEC system with a one-by-one access \cite{5} that is revealed to provide the better energy efficiency than the orthogonal access in MEC scenario considered in our previous study \cite{6}. We jointly optimize the offloading ratio, bit allocation and offloading scheduling that minimize the total energy consumption of vehicles under a given deadline, whose solutions are verified to significantly reduce the total energy consumption of vehicles compared to the benchmarks via numerical results.

\vspace{-0.3cm}
\section{System Model}
\begin{figure}[!t]
\centerline{\includegraphics[width=3.2in]{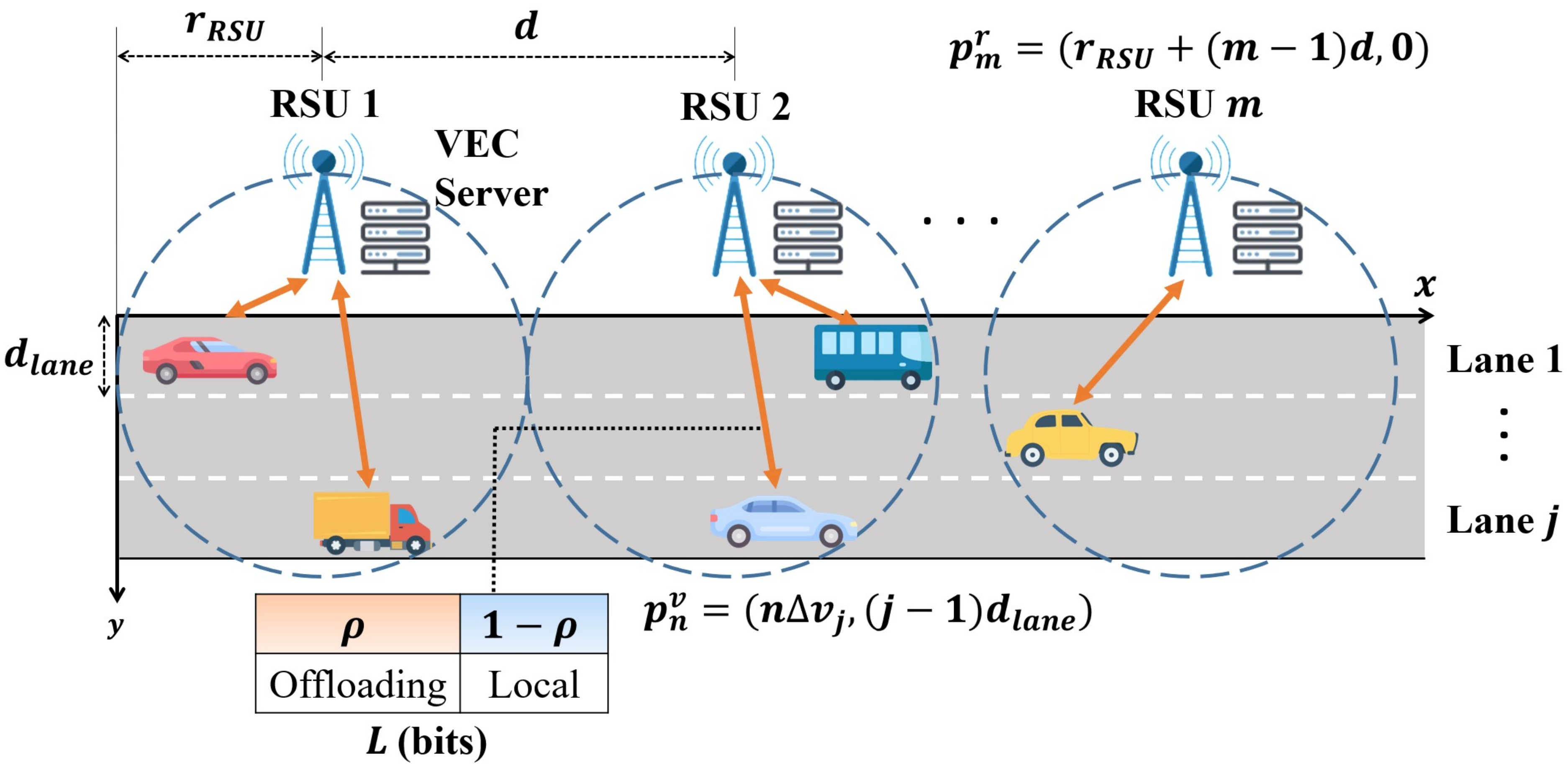}}
\vspace{-0.3cm}
\caption{Illustration of the task offloading in VEC systems.}
\label{fig1}
\vspace{-0.5cm}
\end{figure}

In this paper, we consider a VEC system including $K$ vehicles and $M$ RSUs as shown in Fig.~\ref{fig1}. The RSUs are placed along the unidirectional road with $J$ lanes, the distance between adjacent RSUs is $d$, and the coverage radius of each RSU is $r_ {\text{RSU}}$. We define $\mathcal{M}=\{1,\ldots,M\}$ as the set of RSUs, where the location of RSU $m$ in the xy-plane is calculated as $\mathbf{p}_m^r=(r_{\text{RSU}}+(m-1)d,0)$ for $m\in\mathcal{M}$, with the height $H$. We assume that $K$ vehicles arrive at the first RSU's coverage edge in time $t_k\in\{t_1,\ldots,t_K\}$, the set of which is defined as $\mathcal{K}=\{1,\ldots,K\}$. Also, the vehicles in the same lane have the same velocity, and the velocity of each lane $j$ is assumed to be $v_j\in\{v_1,\ldots,v_J\}$ \cite{7}.\\
\indent Here, we develop the optimal offloading procedure with the aim of minimizing the total energy consumption of all vehicles. To enable the offloading of a given application of each vehicle, the following steps need to be performed. First, the vehicle $k \in \mathcal{K}$ transmits the input data to be computed at the nearest RSU via uplink transmission. Next, the RSU computes the received data. Lastly, the RSU transmits the output of application to vehicle $k$ via downlink transmission. Frequency division duplex (FDD) is assumed, where equal bandwidth $B$ is allocated for both uplink and downlink. Accordingly, there is no interference between uplink and downlink communication. For tractability, the time horizon $T$ is equally divided into $N$ frames as shown in Fig. 2, and each frame duration is $\Delta$ with satisfying $T=N\Delta$. The frame duration $\Delta$ is supposed to be small enough so that the vehicle's position is approximately constant within each frame \cite{8}. Under these circumstances, the position of the vehicle in the $j$th lane on the ground plane at the $n$th frame can be represented as $\mathbf{p}_{j,n}^v=(n\Delta v_j,(j-1)d_{\text{lane}})$, where $j=1,\ldots,J\;\textrm{, }n\in\mathcal{N}=\{1,\ldots,N\}$ and $d_{\text{lane}}$ denotes the lane width. In each frame, the vehicle can communicate with the nearest RSU for offloading. Following \cite{9}, the channel gain between the vehicle $k$ and the adjacent RSU at the $n$th frame is given by $\mathbf{h}_k[n]=\mathbf{h}_k^s[n]\sqrt{h_k^l[n]}$, where $\mathbf{h}_k^s[n]$ is the small-scale fading coefficient which follows Rayleigh distribution with unit variance, and the large-scale fading coefficient $h_k^l[n]$ to reflect the path-loss is expressed as $h_k^l[n]=h_0/(\lVert\mathbf{p}_{j,n}^v-\mathbf{p}_{m_{\min}}^r\rVert^2+H^2)^{\frac{\alpha}{2}}$ with $\Vert\cdot\Vert$ being the norm-2 function, $m_{\min}\in\mathcal{M}$ being the index of the closest RSU, $\alpha$ being the path loss exponent, and $h_0$ being the received power at the reference distance $d=1$m for a transmission power of $1$W. We assume that the channel noise is an additive white Gaussian with zero mean and power spectral density $N_0$ [dBm/Hz].\\
\indent In this paper, we adopt an one-by-one access introduced in \cite{5}, a simple but powerful scheduling method, where only one vehicle can be served at each frame (c.f., Fig. 2(b)). Compared to a conventional orthogonal multiple access (c.f., Fig. 2(a)), where one frame is equally divided and assigned to all vehicles so that each vehicle can communicate with RSU within a single time slot of duration $\delta=\Delta/K$ per each frame, the one-by-one access scheme is verified to be superior by numerical results in Sec. V. This is because the remaining vehicles except the closest vehicle to RSU keep "mute", which can save the energy. The further discussions are shown in the later part of this paper. To adopt the one-by-one scheduling mechanism, the time-varying wake-up scheduling variables are defined as $\{a_k^q[n]\}_{n=1}^{N-2}$, where $q=u$ and $q=d$ stand for uplink and downlink, respectively. If $a_k^q[n]=1$, the vehicle $k$ offloads to the nearest RSU, otherwise $a_k^q[n]=0$. The task of the vehicle $k\in\mathcal{K}$ can be quantified with the number $L_k$ of input bits, the number $C_k$ of CPU cycles per input bit for computation, and the number $\kappa_k$ of output bits produced by computation per input bit. The size of input data is in general much larger than the output data, i.e., $0<\kappa_k<1$.  All tasks offloaded by vehicles need to be computed within the deadline $T$ for the completion.\\
\begin{figure}[!t]
\centerline{\includegraphics[width=3.2in]{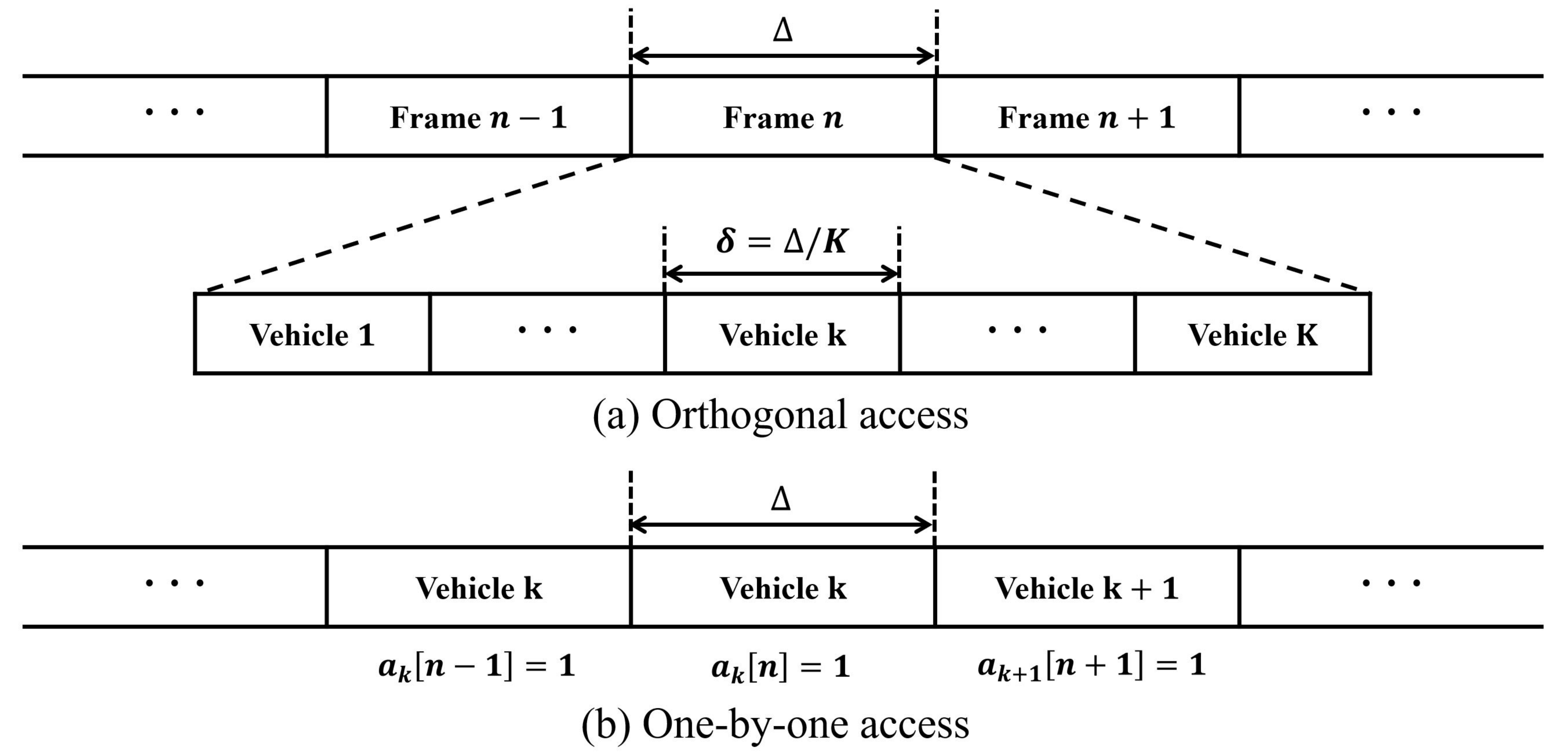}}
\vspace{-0.3cm}
\caption{Frame structure of the considered VEC system: (a) orthogonal access \cite{6}, (b) one-by-one access.}
\label{fig2}
\vspace{-0.5cm}
\end{figure}
\indent For offloading procedure, the computation energy consumption to execute the application of vehicle $k$ with $l$ input bits when the CPU operates at frequency $f$ is calculated by 
\vspace{-0.1cm}
\begin{equation}
E_k^c(l,f)=\gamma C_klf^2, \label{eq5}
\end{equation}
where $f$ [CPU cycles/s] represents the operating frequency of the processor, and $\gamma$ denotes the effective switched capacitance of the processor related to the chip architecture \cite{10}.\\
\indent According to standard information-theoretic arguments, when the vehicle $k$ transmits $l_k^u[n]$ bits in uplink during the $n$th frame of duration $\Delta$, the communication energy consumption of the vehicle $k$ in one-by-one access scheme is given as
\vspace{-0.1cm}
\begin{equation}
E_{k,n}^{\text{one}}(a_k^u[n],l_k^u[n])=a_k^u[n]\frac{N_0B\Delta}{\lVert \mathbf{h}_k[n] \rVert ^2}\bigg(2^{\frac{l_k^u[n]}{B\Delta}}-1\bigg). \label{eq7}
\end{equation}
It is noticed in \eqref{eq7} that the communication energy consumption depends on the scheduling variables $a_k^u[n]$, the number of transmission bits $l_k^u[n]$, and the channel condition $h_k[n]$ affected by the communication distance. As comparison, in orthogonal access \cite{6}, all the vehicles transmit the input data during the allocated slot duration $\delta$ with satisfying $\delta = \Delta/K$ at each frame, which yields the communication energy consumption of the vehicle $k$ at time slot $n$ as
\vspace{-0.1cm}
\begin{equation}
E_{k,n}^{\text{orth}}(l_k^u[n])=\frac{N_0B\delta}{\lVert \mathbf{h}_k[n] \rVert ^2}\bigg(2^{\frac{l_k^u[n]}{B\delta}}-1\bigg). \label{eq8}
\end{equation}

\vspace{-0.3cm}
\section{Energy-efficient Offloading with One-by-one Access}
In this section, we formulate the problem to minimize the total energy consumption of $K$ vehicles by jointly optimizing the bit allocation, offloading ratio and scheduling for one-by-one access scheme. For reference, the total energy consumption of vehicles in local execution and offloading case with the orthogonal access \cite{6} are briefly discussed first.

\vspace{-0.3cm}
\subsection{Energy Consumption for Local Execution}
In this part, we consider the total energy consumption of overall vehicles when all the applications are processed locally. In order to process $L_k$ bits within $T$ seconds, the CPU frequency of vehicle $k$ needs to be selected as $f_k^v=C_kL_k/T$.
According to \eqref{eq5}, the total energy consumption for local execution is obtained by $\sum_{k=1}^{K} E_k^{\text{local}}(L_k)=\sum_{k=1}^{K} \gamma_k^vC_k^3L_k^3/T^2$, where $\gamma_k^v$ is the effective switched capacitance of the vehicle $k$'s processor.

\vspace{-0.3cm}
\subsection{Minimal Energy Consumption for Orthogonal Multiple Access (Our Previous Work \cite{6})}
In our previous work \cite{6}, an orthogonal access scheme is developed for VEC systems to minimize the total energy consumption of vehicles. To this end, the joint optimization of bit allocation and offloading ratio between local execution and RSU execution is studied, where the vehicle $k$ offloads the ratio $\rho_k$ of the input bits to the RSU and locally computes the remaining portion $(1- \rho_k)$ of the input bits. At the $n$th frame, we denote $l_k^u[n]$ as the number of uplink bits transmitted from the vehicle $k$ to the RSU, $l_k^c[n]$ as the number of bits computed for the task of the vehicle $k$ at the RSU, and $l_k^d[n]$ as the number of downlink bits transmitted from the RSU to the vehicle $k$. Since, in the orthogonal access, the offloading process is analyzed in frame-by-frame manner \cite{6}, \cite{8}, the energy consumption of the vehicle is expressed as $E^{\text{orth}}_{\text{total}}(l_k^u[n], \rho_k)\!=\!\!\sum_{k=1}^{K}\sum_{n=1}^{N-2}E_{k,n}^{\text{orth}}(l_k^u[n])+\sum_{k=1}^{K}E_k^{\text{local}}((1-\rho_k)L_k)$. To minimize the total energy consumption of the vehicle, the joint optimization problem of bit allocation and offloading ratio can be formulated as (8) in \cite{6}, whose solutions are detailed in \cite{6}.

\vspace{-0.3cm}
\subsection{Minimal Energy Consumption for One-by-one Access}
We now formulate the optimization problem when adopting the one-by-one access scheme \cite{5} for VEC systems, and then propose Algorithm 1 to resolve the formulated problem. Herein, we consider not only the bit allocation and the offloading ratio between local execution and RSU execution, but also the scheduling variables for one-by-one access. To this end, the total energy consumption of vehicles is given by $E^{\text{one}}_{\text{total}}(l_k^u[n], \rho_k, a_k^u[n])=\sum_{k=1}^{K}\sum_{n=1}^{N-2}E_{k,n}^{\text{one}}(a_k^u[n],l_k^u[n])+\sum_{k=1}^{K}E_k^{\text{local}}((1-\rho_k)L_k)$. Let us denote $\mathcal{Z}=\{l_k^u[n],l_k^c[n],l_k^d[n],a_k^u[n],a_k^d[n],\rho_k\}$ as the set of optimization variables, and therefore the optimization problem for one-by-one access is formulated as
\vspace{-0.1cm}
\begin{subequations} \label{eq14}
\begin{align}
& \underset{\mathcal{Z}}{\text{minimize}} \;\;E^{\text{one}}_{\text{total}}(\mathcal{Z}) \label{eq14a} \\
& \text{s.t.} \;\frac{l_k^u[n]}{B\Delta} \leq a_k^u[n] \log_2\Bigg(1+\frac{P_{\textrm{max}}\lVert h_k[n] \rVert ^2}{N_0B}\Bigg), \; \forall k, n \in \tilde{\mathcal{N}}, \label{eq14b} \\
& \;\;\;\;\; \frac{l_k^d[n+2]}{B\Delta} \leq a_k^d[n+2] \log_2\Bigg(1+\frac{P_{\textrm{RSU}}\lVert h_k[n+2] \rVert ^2}{N_0B}\Bigg),\nonumber \\
& \;\;\;\;\; \forall k, n \in \tilde{\mathcal{N}}, \label{eq14c} \\
& \;\;\;\;\; \sum_{i=1}^{n}l_k^c[i+1] \leq \sum_{i=1}^{n}l_k^u[i], \; \forall k, n \in \tilde{\mathcal{N}}, \label{eq14d} \\
& \;\;\;\;\;\sum_{i=1}^{n}l_k^d[i+2] \leq \kappa_k\sum_{i=1}^{n}l_k^c[i+1], \; \forall k, n \in \tilde{\mathcal{N}}, \label{eq14e} \\
& \;\;\;\;\; \sum_{k=1}^{K} a_k^u[n]=1,\; \sum_{k=1}^{K} a_k^d[n+2]=1, \; \forall k, n \in \tilde{\mathcal{N}}, \label{eq14f} \\
& \;\;\;\;\; \sum_{n=1}^{N-2}l_k^u[n] = \rho_kL_k,\; \sum_{n=1}^{N-2}l_k^c[n+1] = \rho_kL_k, \; \forall k, \label{eq14g} \\
& \;\;\;\;\; \sum_{n=1}^{N-2}l_k^d[n+2] = \kappa_k \rho_kL_k, \; \forall k, \label{eq14h} \\
& \;\;\;\;\; 0 \leq \rho_k \leq 1,  \; \forall k, \label{eq14i} \\
& \;\;\;\;\; a_k^u[n],a_k^d[n]\in\{0,1\}, \; \forall k, n \in \mathcal{N}, \label{eq14j} \\
& \;\;\;\;\; l_k^u[n],\:l_k^c[n],\:l_k^d[n] \geq 0, \; \forall k, n \in \mathcal{N}, \label{eq14k}
\end{align}
\end{subequations}
where the constraints \eqref{eq14b} and \eqref{eq14c} guarantee that the achievable rates in uplink and downlink are larger than or equal to the number of transmitted bits in the corresponding links. Also, an equality constraint \eqref{eq14f} is for scheduling of one-by-one access to satisfy that the only one vehicle can communicate with RSU in each frame.\\
\begin{figure*}[!t]
\begin{align}
&\mathcal{L}(\mathcal{Z},\mathcal{Y})=\sum_{k=1}^{K}\sum_{n=1}^{N-2}a_k^u[n]F_k^u[n]+\sum_{k=1}^{K}\sum_{n=1}^{N-2}a_k^d[n+2]F_k^d[n]+\sum_{k=1}^{K} \frac{\gamma_k^c{C_k^3}}{T^2}(1-\rho_k)^3L_k^3+\sum_{k=1}^{K}\sum_{n=1}^{N-2}\lambda_k^u[n]\frac{l_k^u[n]}{B\Delta} \nonumber \\
&+\sum_{k=1}^{K}\sum_{n=1}^{N-2}\lambda_k^d[n]\frac{l_k^d[n+2]}{B\Delta}-\sum_{k=1}^{K}\sum_{n=1}^{N-2}\mu_k^u[n]l_k^u[n]+\sum_{k=1}^{K}\sum_{n=1}^{N-2}(\mu_k^u[n]-\kappa_k\mu_k^d[n])l_k^c[n+1]+\sum_{k=1}^{K}\sum_{n=1}^{N-2}\mu_k^d[n]l_k^d[n+2] \nonumber \\
&+\sum_{k=1}^{K}u_k^u\bigg(\sum_{n=1}^{N-2}l_k^u[n]-\rho_kL_k\bigg)+\sum_{k=1}^{K}u_k^c\bigg(\sum_{n=1}^{N-2}l_k^c[n+1]-\rho_kL_k\bigg)+\sum_{k=1}^{K}u_k^d\bigg(\sum_{n=1}^{N-2}l_k^d[n+2]-\kappa_k\rho_kL_k\bigg), \label{eq15}
\end{align}
\hrulefill
\vspace{-0.5cm}
\end{figure*}
\indent Since the problem \eqref{eq14} is non-convex and mixed-integer optimization problem which cannot be directly solved by using standard convex optimization techniques. To address the non-convexity, we adopt the corresponding Lagrange dual problem of \eqref{eq14}. Let us define $\mathcal{Y}=\{\lambda_k^u[n],\lambda_k^d[n],\mu_k^u[n],\mu_k^d[n],u_k^u,u_k^c,u_k^d\}$ as the set of Lagrange dual variables corresponding to \eqref{eq14b}-\eqref{eq14e}, \eqref{eq14g} and \eqref{eq14h}, respectively. Then, the Lagrangian of problem \eqref{eq14} is defined as \eqref{eq15}, where $F_k^u[n]$ and $F_k^d[n]$ is expressed as
\vspace{-0.1cm}
\begin{equation}
F_k^u[n]\!=\!\frac{N_0B\Delta}{\lVert \mathbf{h}_k[n] \rVert ^2}\bigg(\!2^{\frac{l_k^u[n]}{B\Delta}}-1\!\bigg)\!-
\lambda_k^u[n]\log_2\!\bigg(\!1+\frac{P_{\textrm{max}}\lVert h_k[n] \rVert ^2}{N_0B}\!\bigg), \label{eq16}
\end{equation}
and
\vspace{-0.1cm}
\begin{equation}
F_k^d[n]=-\lambda_k^d[n]\log_2\bigg(1+\frac{P_{\textrm{RSU}}\lVert h_k[n+2] \rVert ^2}{N_0B}\bigg). \label{eq17}
\end{equation}
Given $\mathcal{Y}$, the optimal offloading ratio $\rho_k^{\text{opt}}$ can be obtained by applying Karush-Kuhn-Tucker (KKT) conditions. The Lagrange dual function of problem \eqref{eq14} is given by
\vspace{-0.1cm}
\begin{equation}
g(\mathcal{Y}) = \left\{ \begin{array}{ll}
\underset{\mathcal{Z}}\min\; \mathcal{L}(\mathcal{Z},\mathcal{Y}) \\
\textrm{s.t.} \;\; \eqref{eq14f}, \eqref{eq14i}-\eqref{eq14k}. \label{eq18} 
\end{array} \right.
\end{equation}
In order to minimize dual function $\mathcal{L}(\mathcal{Z},\mathcal{Y})$, the stationary point $\rho_k^{\textrm{opt}}$ to make the derivative of $\mathcal{L}$ with respect to $\rho_k$ equal to zero can be obtained as
\vspace{-0.1cm}
\begin{equation}
    \rho_k^{\textrm{opt}}=1-\sqrt{\bigg[\frac{(u_k^u+u_k^c+\kappa_k u_k^d) T^2}{3\gamma_k^c{C_k^3}L_k^2}\bigg]_0^1}, \label{eq20} 
\end{equation}
where $[c]^b_a=\textrm{min}\{\textrm{max}\{a,c\},b\}$.
In a similar way, to minimize $F_k^u[n]$ and $F_k^d[n]$, the optimal scheduling variables $a_k^{u,opt}[n]$ and $a_k^{d,opt}[n]$ for $k \in K$ and $n \in N$ are calculated by the following theorem.
\newtheorem{Theorem}{Theorem}
\begin{Theorem}
Given $\mathcal{Y}$ and $l_k^u[n]$ for all $k\in\mathcal{K}$ and $n\in\mathcal{\tilde{N}}$, the optimal scheduling variables for Lagrange dual function are obtained as
\vspace{-0.1cm}
\begin{subequations} \label{eq21}
\begin{align}
& a_k^{u,\textrm{opt}}[n] = \left\{ \begin{array}{ll}
1 & k=\arg \underset{k'\in\mathcal{K}}\min F_{k'}^{u}[n]\\
0 & \textrm{otherwise,} \label{eq21a} \\
\end{array} \right. \\
& a_k^{d,\textrm{opt}}[n] = \left\{ \begin{array}{ll}
1 & k=\arg \underset{k'\in\mathcal{K}}\min F_{k'}^d[n]\\
0 & \textrm{otherwise.} \label{eq21b} \\
\end{array} \right. 
\end{align}
\end{subequations}
\end{Theorem}
As a result, the scheduling variables $a_k^u[n]$ and $a_k^d[n]$ should be chosen so as to minimize $F_k^u[n]$ and $F_k^d[n]$ under the constraints \eqref{eq14f} and \eqref{eq14j}. Given $a_k^{u,\textrm{opt}}[n]$, $a_k^{d,\textrm{opt}}[n]$ and $\rho_k^{\textrm{opt}}$, the optimization problem \eqref{eq14} is simplified as
\vspace{-0.1cm}
\begin{subequations} \label{eq22}
\begin{align}
& \underset{\{l_k^u[n]\},\{l_k^c[n]\},\{l_k^d[n]\}}{\text{minimize}} \sum_{k=1}^{K}\sum_{n=1}^{N-2}E_{k,n}^u(l_k^u[n]) \label{eq22a} \\
& \text{s.t.} \;\eqref{eq14b}-\eqref{eq14e},\eqref{eq14g}-\eqref{eq14i},\eqref{eq14k}. \label{eq22b} 
\end{align}
\end{subequations}
Since the problem \eqref{eq22} is convex, we can solve this problem using standard convex optimization solver such as CVX \cite{11}.
After that, we can solve the dual problem as follows:
\vspace{-0.1cm}
\begin{subequations} \label{eq23}
\begin{align}
& \underset{\mathcal{Y}}{\text{max}} \;\;g(\mathcal{Y}) \label{eq23a} \\
& \;\;\text{s.t.} \;\;\lambda_k^u[n],\lambda_k^d[n],\mu_k^u[n],\mu_k^d[n]\ge0, \label{eq23b} 
\end{align}
\end{subequations}
Since the dual problem \eqref{eq23} is concave with respect to $\mathcal{Y}$, the subgradient method is adopted so that converging the global point can be guaranteed \cite{12}. Accordingly, the dual variables in each iteration are given by
\vspace{-0.1cm}
\begin{align}
\lambda_k^{u,z+1}[n]&=\bigg[\lambda_k^{u,z}[n]+\pi_1\bigg(\frac{l_k^u[n]}{B\Delta} \nonumber \\
&\;\;\;\;\;\;-a_k^u[n]\log_2\Big(1+\frac{E_k^u(L_{k,n}^u)h_{k,n}}{N_0B\Delta}\Big)\bigg)\bigg]^+, \label{eq24} \\
\lambda_k^{d,z+1}[n]&=\bigg[\lambda_k^{d,z}[n]+\pi_2\bigg(\frac{l_k^d[n]}{B\Delta} \nonumber \\
&\;\;\;\;\;\;-a_k^d[n]\log_2\Big(1+\frac{E_k^d(L_{k,n}^u)h_{k,n}}{N_0B\Delta}\Big)\bigg)\bigg]^+, \label{eq25} \\
\mu_k^{u,z+1}[n]&= \big[\mu_k^{u,z}[n]+\pi_3(l_k^c[n+1]-l_k^u[n])\big]^+, \label{eq26} \\
\mu_k^{d,z+1}[n]&= \big[\mu_k^{d,z}[n]+\pi_4(l_k^d[n+2]-\kappa_kl_k^c[n+1])\big]^+, \label{eq27} \\
u_k^{u,z+1}&=u_k^{u,z}+\pi_5\bigg(\sum_{n=1}^{N}l_{k,n}^u-\rho_kL_k\bigg), \label{eq28} \\
u_k^{c,z+1}&=u_k^{c,z}+\pi_6\bigg(\sum_{n=1}^{N}l_{k,n}^c-\rho_kL_k\bigg), \label{eq29} \\
u_k^{d,z+1}&=u_k^{d,z}+\pi_7\bigg(\sum_{n=1}^{N}l_{k,n}^d-\kappa_k\rho_kL_k\bigg), \label{eq30}
\end{align}
where the superscript 'z' represents the iteration index, $[c]^+=\textrm{max}\{0,c\}$ and $\{\pi_i\}_{i=1}^7$ are step sizes. The overall process is shown in Algorithm 1.
\renewcommand{\algorithmicrequire}{\textbf{Input:}}
\renewcommand{\algorithmicensure}{\textbf{Output:}}
\begin{algorithm}[!t]
\caption{Joint optimization of bit allocation, offloading ratio and scheduling for one-by-one access in VEC systems.}
\begin{algorithmic}[1]
\State{\textbf{Initialize} $\mathcal{Y}$ and $l_k^u[n]$, $\forall k$, $n \in \mathcal{N}$.}
\State{\textbf{Repeat}}
\State \, Obtain $\rho_k^{\textrm{opt}}$ using \eqref{eq20}.
\State \, Obtain $a_k^{u,\textrm{opt}}[n]$ and $a_k^{d,\textrm{opt}}[n]$ using \eqref{eq21a} and \eqref{eq21b}.
\State \, Obtain $l_k^{u,\text{opt}}[n]$, $l_k^{c,\text{opt}}[n+1]$ and $l_k^{d,\text{opt}}[n+2]$ for given $\rho_k^{\textrm{opt}}$, $a_k^{u,\textrm{opt}}[n]$ and $a_k^{d,\textrm{opt}}[n]$ using CVX.
\State \, Update $\mathcal{Y}$ with \eqref{eq24}-\eqref{eq30}.
\State{\textbf{Until} convergence.}
\Ensure $\mathcal{Z}^{\text{opt}}$, $\forall k$, $n \in \mathcal{N}$.
\end{algorithmic}
\label{al1}
\vspace{-0.1cm}
\end{algorithm}

\vspace{-0.3cm}
\section{Simulation Results}
In this section, the simulation results are presented to evaluate the performance of our proposed Algorithm 1 in VEC systems. For simulations, we consider the VEC system including a one-way road with three lanes. Each vehicle randomly arrives at the starting point, and it is assumed that the task deadline of all vehicles is equal to $T$. Also, Rayleigh fading is considered for small-scale fading, and 3GPP path loss model \cite{13} is used for large-scale fading. The remaining parameters are shown in Table 1. As a benchmark, the three schemes are considered such as (i) local execution scheme, where all tasks are computed locally, (ii) orthogonal access scheme \cite{6}, where the same time slot is allocated to each vehicle, (iii) one-by-one access scheme with an equal bit allocation that transmits the same number of bits to the uplink and downlink in each frame, without optimizing the bit allocation.\\
\begin{table}[!t]
\renewcommand{\arraystretch}{1}
\vspace{-0.2cm}
\caption{Simulation Parameters}
\vspace{-0.4cm}
\begin{center}
\begin{tabular}{cccc}
\hline\hline
Parameter & Value & Parameter & Value\\
\hline
$r_{\text{RSU}}$ & 250m & $d, d_{\text{lane}}$ & 500m, 4m \\
$H$ & 20m & $\Delta$ & 30ms \\ 
$B$ & 20MHz & $L_k$ & 20-75Mbits \\
$C_k$  & 1550.7 \cite{10} & $\gamma^r, \gamma_k^v$ & $10^{-28}$ \cite{10} \\
$\kappa_k$ & 0.5 & $N_0$ & -114dBm/Hz \\
$v_j$ & 30 - 35m/s & $P_{\textrm{max}}, P_{\textrm{RSU}}$ & 1W, 2W \cite{14} \\
\hline\hline
\end{tabular}
\end{center}
\label{table1}
\vspace{-0.7cm}
\end{table}
\begin{figure}[!t]
\centerline{\includegraphics[width=3.2in]{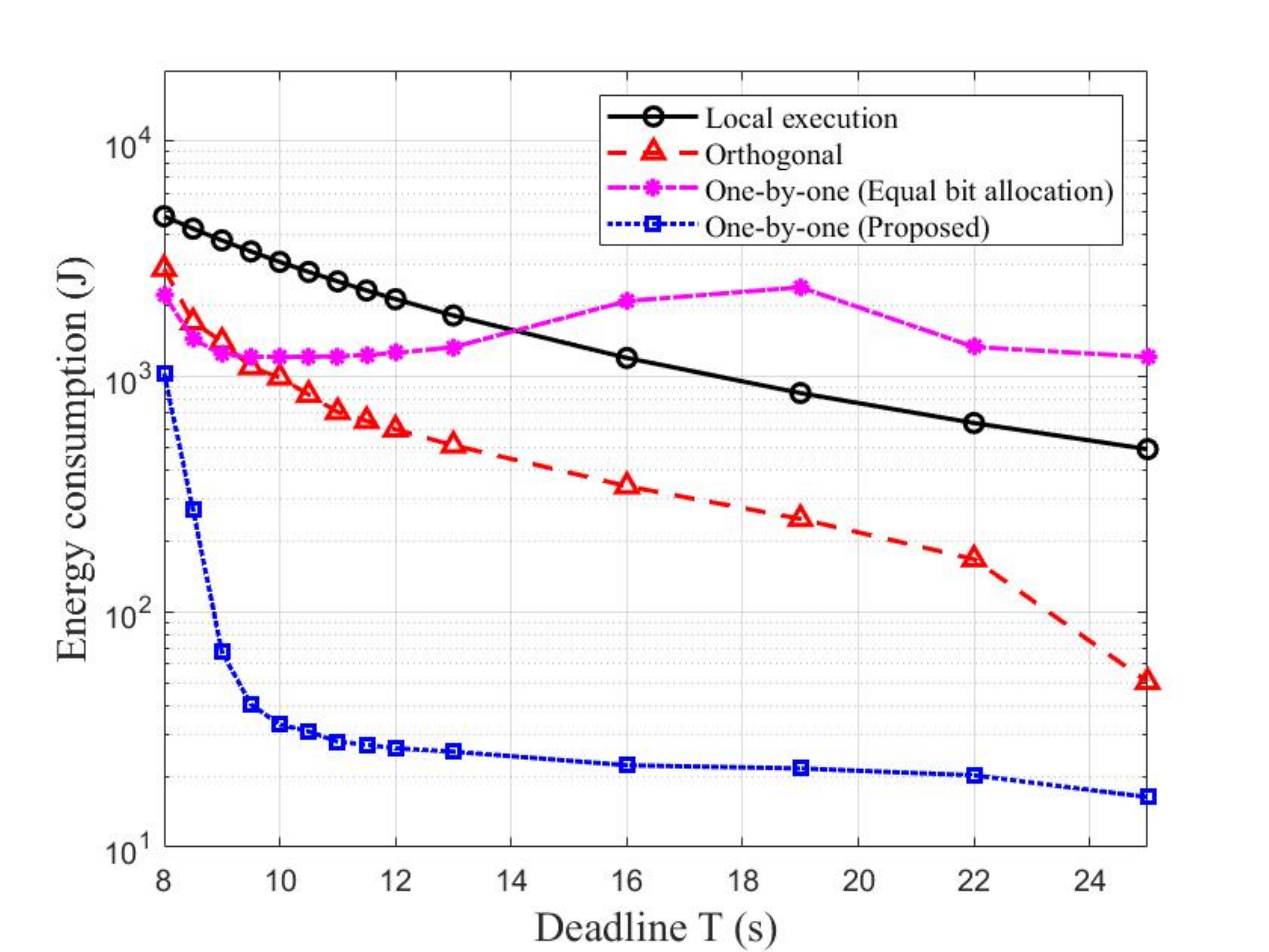}}
\vspace{-0.3cm}
\caption{Total energy consumption of vehicles versus deadline $T$}
\label{fig3}
\vspace{-0.5cm}
\end{figure}
\indent Fig. 3 shows the total energy consumption of vehicles as a function of deadline $T$ on a logarithmic scale, where the number of vehicles is set to 3, and the input bits of each vehicle are set to 75Mbits. It is observed that the proposed one-by-one access scheme consumes the least energy compared to other benchmarks. Furthermore, energy consumption of one-by-one access dramatically decreases around 10s, as it approaches the first RSU. This is because as the distance between the RSU and the vehicle gets closer, the vehicle can offload the larger amount of tasks to the RSU so as to consume the less communication energy. On the other hand, in orthogonal access, where the frame duration is divided and equally allocated for each vehicle, although the vehicle is close to the RSU, it cannot fully utilize the entire duration, yielding the higher energy consumption than one-by-one access scheme. Similarly, in the case of one-by-one with equal bit allocation, since the same amount of bits needs to be offloaded in each frame regardless of channel conditions, the vehicular energy consumption decreases when the vehicle approaches the RSU, and increases again when the vehicle moves away from the RSU. Additionally, at the deadline larger than 14s, the vehicular energy consumption is even higher than that of local execution, which shows that the optimal bit allocation plays an important role in the aspect of energy efficiency.\\
\indent In Fig. 4, we compare the total energy consumption as a function of the number of input bits, where $K=3$ and $T=25s$. In local execution, since the computation energy consumption is proportional to the cube of the input bits, the energy consumption increases significantly as the number of input bits increases. Both orthogonal and one-by-one access scheme are designed to offload the most of tasks, resulting in the sizable energy reduction compared to local execution. In particular, in the case of one-by-one access, where the entire frame duration can be allocated to the vehicle closest to the RSU, it is robust against the large-scale input bits.\\
\indent Fig. 5 compares the total energy consumption as a function of the number of vehicles under $T=25s$. In local execution, the vehicular energy consumption is highest as the number of vehicles increases, since the total energy consumption of vehicles is obtained by summing all the computational energy consumption of each vehicle. Also, in orthogonal access scheme, the available time duration at each vehicle becomes shorter with the increase on the number of vehicles, which results in the larger energy consumption. On the other hand, the one-by-one access scheme achieves the lowest energy consumption as the number of vehicles increases.
\begin{figure}[!t]
\centerline{\includegraphics[width=3.2in]{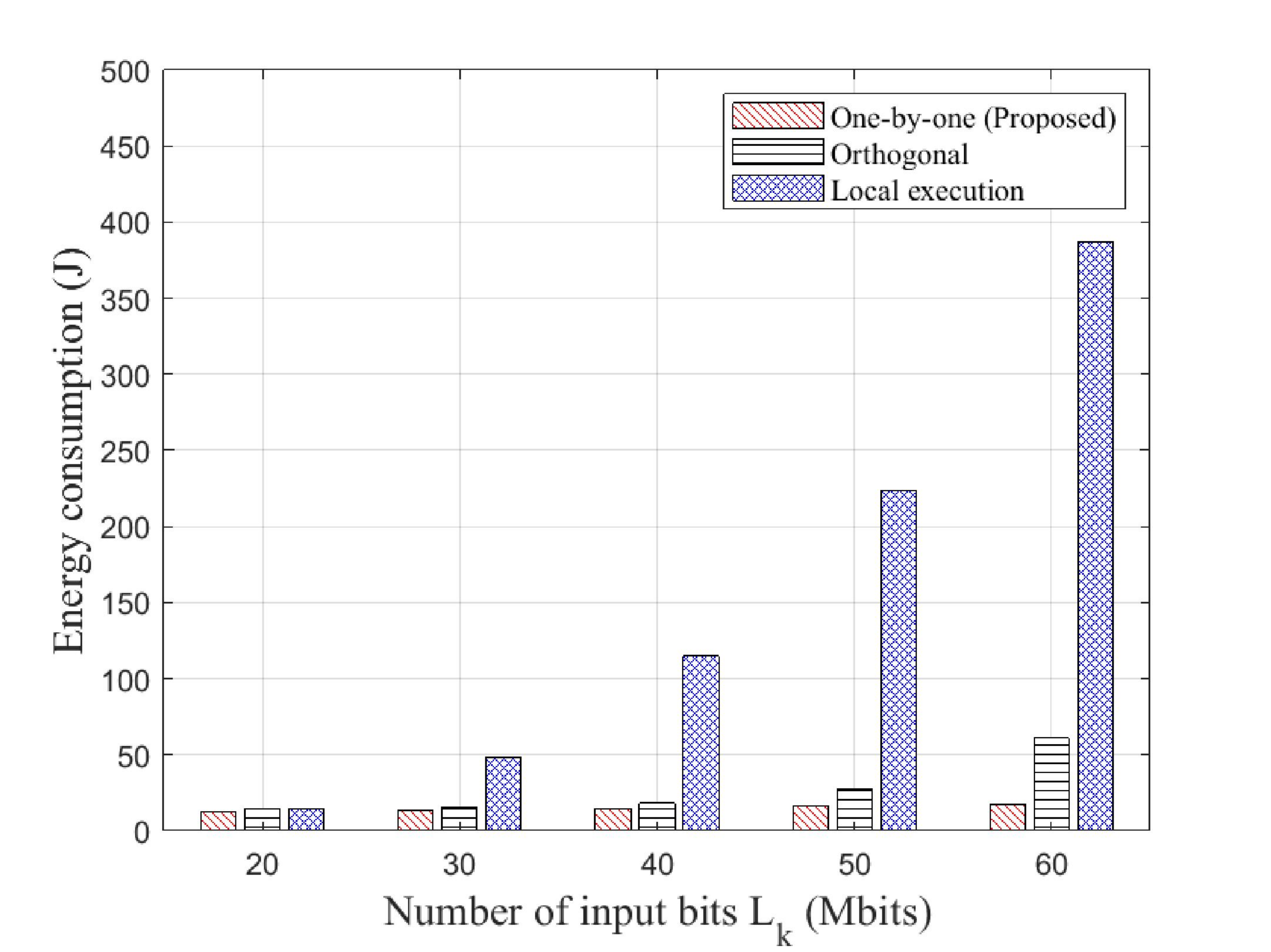}}
\vspace{-0.3cm}
\caption{Total energy consumption of vehicles versus number of input bits $L_k$}
\label{fig4}
\vspace{-0.5cm}
\end{figure}

\vspace{-0.3cm}
\section{Conclusion}
In this paper, an energy-efficient task offloading scheme for VEC system with one-by-one access is proposed. To minimize the total energy consumption of vehicles, we jointly optimize the offloading ratio, bit allocation, and offloading scheduling under a given deadline. The non-convex and mixed-integer optimization problem is formulated and solved by adopting Lagrange dual problem. Via simulations, we verify that the proposed energy-efficient offloading scheme can significantly reduce the total energy consumption of vehicles compared to the benchmarks. As a future work, a scenario considering traditional non-orthogonal multiple access (NOMA) and rate-splitting multiple access (RSMA) can be studied.

\begin{figure}[!t]
\centerline{\includegraphics[width=3.2in]{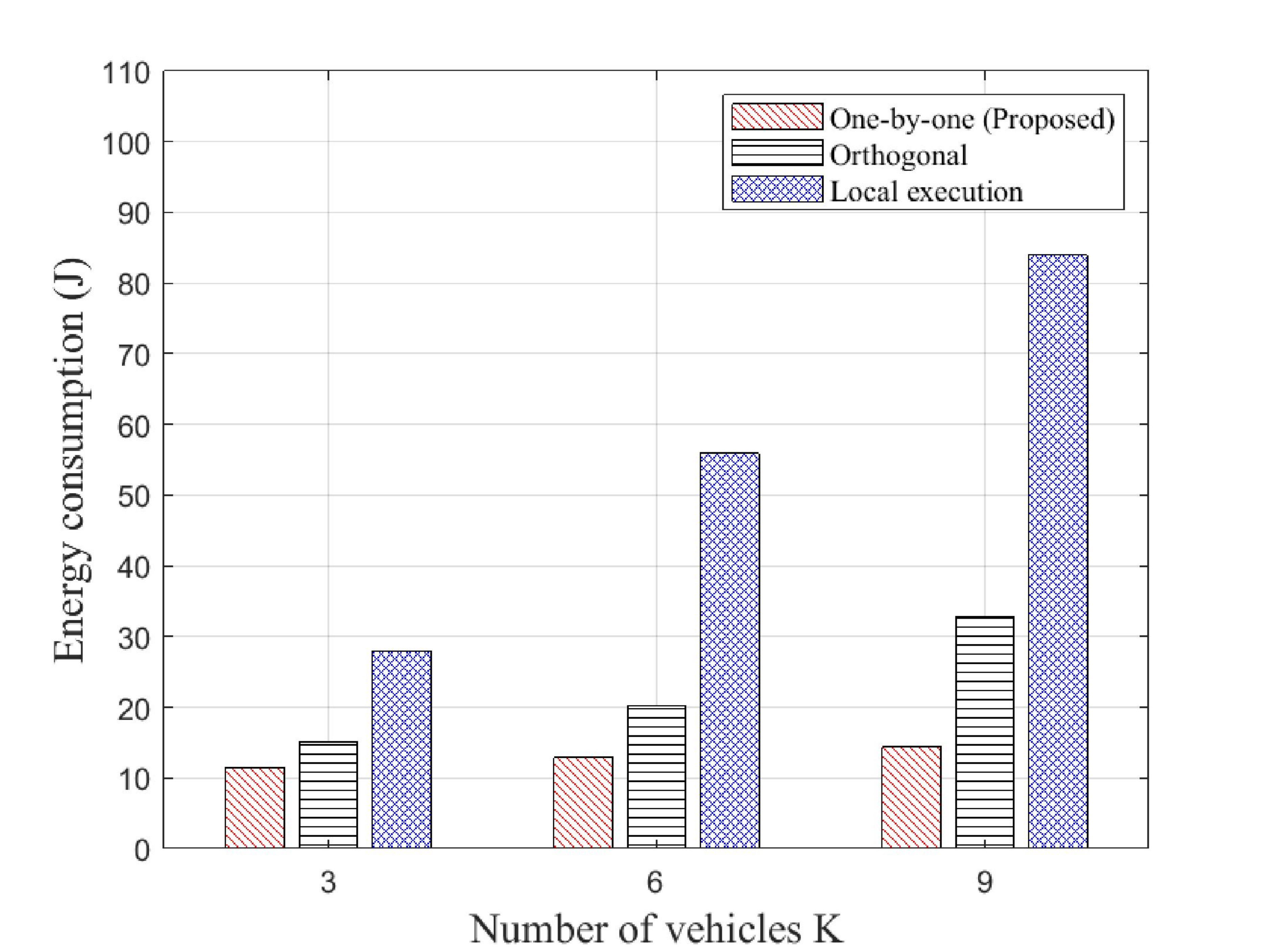}}
\vspace{-0.3cm}
\caption{Total energy consumption of vehicles versus number of vehicles $K$}
\label{fig5}
\vspace{-0.5cm}
\end{figure}


\ifCLASSOPTIONcaptionsoff
  \newpage
\fi


\end{document}